\documentclass[format=sigplan]{acmart}
\makeatletter
\renewcommand\footnotetextcopyrightpermission[1]{}
\makeatother
\usepackage{listings}

\acmPrice{}
\acmDOI{}
\acmISBN{}
\begin{document}
\setcopyright{none}
\title{Extended abstract: Type oriented programming for task based parallelism}
	
\author{Nick Brown}
\affiliation{EPCC, James Clerk Maxwell Building, Peter Guthrie Tait Road, Edinburgh}
\email{nick.brown@ed.ac.uk}
\author{Ludovic Capelli}
\affiliation{EPCC, James Clerk Maxwell Building, Peter Guthrie Tait Road, Edinburgh}
\author{James Mark Bull}
\affiliation{EPCC, James Clerk Maxwell Building, Peter Guthrie Tait Road, Edinburgh}

\acmConference[TyDe 2017]{2nd Workshop on Type-Driven Development}{3rd September, 2017}{Oxford, UK}

\maketitle
	
\section{Challenge}
Writing parallel codes is difficult \cite{Skillicorn} and exhibits a fundamental trade-off between abstraction and performance. The high level language abstractions designed to simplify the complexities of parallelism make certain assumptions that impacts performance and scalability. On the other hand lower level languages, providing many opportunities for optimisation, require in-depth knowledge and the programmer to consider tricky details of parallelism. An approach is required which can bridge the gap and provide both the ease of programming and opportunities for control and optimisation.

\section{Type oriented programming}
By optionally decorating their codes with additional type information, programmers can either direct the compiler to make certain decisions or rely on sensible default choices. In \cite{mesham} we introduced a research based programming language, Mesham, which explores these ideas in relation to data parallelism. This is illustrated in listing \ref{lst:first} with three declarations of variables.

\begin{lstlisting}[frame=lines,caption={Type examples in the Mesham language},label={lst:first}]
var a:Int;
var b:Int :: allocated[single[on[0]]];
var c:Int :: allocated[single[on[0]]] :: channel[0,1];
\end{lstlisting}

The first declaration determines that \emph{a} is an integer and, in the absence of further type information, the compiler generates code that allocates this variable on every process. The behaviour of an assignment \emph{a:=22} is to perform a local assignment on every process. In the second declaration we have added extra type information to further direct the compiler, in this case explicitly specifying that \emph{b} will be allocated only on process 0. Based upon this additional information the same assignment \emph{b:=22} generates a local assignment on process 0 and remote communications on every other process. By default the behaviour of these communications is simple and safe, however might not be particularly performant. In declaration three the programmer has added extra type information to guide the compiler to handle remote data access as a point to point communication (a channel) rather than the default RMA. In our approach the programmer, after experimentation and profiling, can further directed parallelism in a high level manner without having to consider the low level implementation details or significantly rewrite their code. The use of types in listing \ref{lst:first} is different from simply specifying variable storage because, for instance, the assignment \emph{a:=b} invokes a broadcast from process 0 of the value held in \emph{b} to all other processes, writing this into their local \emph{a}. 

We denote this combination of types, which then determines the behaviour of all variable usage, using the \emph{::} operator. This is known as a type chain. Precedence is from right to left, so certain types can override the behaviour of other types based upon their order in the chain. Types can be arbitrarily chained together and any potentially conflicting combinations are handled by this precedence rule. The specific types themselves are separate from the core semantics of the language which means that, from a language perspective, it is trivial to add or remove types for specific domains. This approach has been applied to areas from traditional HPC \cite{easc} to graph based codes \cite{graph}.

Our use of types is different from annotation approaches, such as OpenMP \cite{openmp}, because types are part of, and integrate fully with, the language rather than a bolt on. Therefore the programmer has flexibility to create new types in their code and reasoning about type information using existing language constructs. Through constructing type chains we provide a mechanism for building up complex type information in a structured, hierarchical, manner and it is this type chain that provides the behaviour of operations performed on the variable throughout its life.

\section{Task based parallelism}
Many traditional HPC codes have been oriented around data parallelism, where a data is split up and distributed amongst processes. However associated techniques such as halo swapping often result in a bulk synchronous style of parallelism, where processes proceed in computation and then communication/synchronisation steps. This synchronisation, which is often global, is inefficient and to reach the scale of millions of processes must be avoided. 

In task based parallelism computational tasks themselves are decomposed amongst the processes. Typically tasks are scheduled based upon a number of dependencies and will execute once these dependencies are met. Task based parallelism forces the programmer to break away from their bulk synchronous approach and promotes the asynchronous nature of codes. Whilst there is active research into task based programming models, the level of abstraction is again a major challenge. Many existing technologies rely on the runtime to make sensible decisions, such as scheduling. This can have a significant impact on performance without the user being able to explicitly control or tune important parameters.

\section{Types for task based parallelism}
Our use of types so far followed the data parallelism approach but we believe that, by decorating functions, types can also address the abstraction challenge of task based parallelism. In Mesham the declaration of a function, \emph{myFunction} which takes in and returns an integer follows \emph{function Int myFunction(var a:Int)}. When called, the behaviour is to execute the function immediately and return the value. It is possible to apply addition type information to the function declaration. The addition of the \emph{spawnable} type such as \emph{function Int myFunction(var a:Int):spawnable} overrides this default behaviour. Because of the additional type, calls to the function will schedule it for execution on a thread rather than executing it directly. The \emph{spawnable} type effectively transforms functions into tasks, one per thread run concurrently.

An important aspect of scheduling these tasks is to have some way of referencing them. The semantics of the \emph{spawnable} type is that function calls will return a variable of the type \emph{Future[X]} (where \emph{X} is the actual return type, in this case \emph{Future[Int]}.) This future can be used as a handle to test and wait for completion. Listing \ref{lst:second} illustrates a simple Fibonacci example where the \emph{fib} function is marked as \emph{spawnable} (a concurrent task.) Based upon the function calls at lines 4 and 5, upon execution the task schedules two further recursive \emph{fib} tasks and the calling of these functions immediately return futures to these as the variables \emph{f1} and \emph{f2}. Lines 6 and 7 synchronise on the futures (waits for their corresponding tasks to complete), before adding the integer values together and returning the result.

\begin{lstlisting}[frame=lines,caption={Fibonacci task parallelism with explicit synchronistion on futures},label={lst:second}]
function Int fib(var val:Int) : spawnable {
    if (val == 0 || val == 1) return val;
    var f1,f2 : Future[Int];
    f1:=fib(val-1);
    f2:=fib(val-2);
    synchronise(f1);
    synchronise(f2);
    return f1.val + f2.val;
}
\end{lstlisting}

However the code in listing \ref{lst:second} is naive as the explicit synchronisations block the calling thread which is wasteful. To avoid this we allow the programmer, via type information, to encode the dependencies of tasks within their code. Now the scheduler will not execute the task until these dependencies are met and hence there is no explicit synchronisation or blocking of threads. Listing \ref{lst:third} is the same Fibonacci task parallel code but using task dependencies, via the \emph{dependencies} type, on the \emph{add} function instead of explicit synchronisation. The behaviour of the \emph{dependencies} type is that the decorated function will accept both normal valued arguments (in this case integers for variables \emph{a} and \emph{b}) as well as futures. If futures are provided (as is the case in listing \ref{lst:third}) then the scheduler will wait until the tasks that they depend upon have completed before executing the scheduled, decorated, function. In this manner execution of the \emph{fib} task returns a future on the \emph{add} task which itself is dependent on futures of recursive calls to the \emph{fib} task. 

\begin{lstlisting}[frame=lines,caption={Fibonacci task parallelism with dependencies},label={lst:third}]
function Int fib(var val:Int) : spawnable {
    if (val == 0 || val == 1) return val;
    var f1,f2 : Future[Int];
    f1:=fib(val-1);
    f2:=fib(val-2);
    return add(f1, f2);
}

function Int add(var a:Int, var b:Int) : spawnable :: dependencies {
    return a + b;
}
\end{lstlisting}

It is possible to omit the \emph{spawnable} type whilst keeping the \emph{dependencies} type. In this case the same dependencies behaviour is present, with function execution being immediate and blocking rather than a task. This is how we implement the \emph{synchronise} call of listing \ref{lst:second}.

The idea of programmers decorating functions with type information in order to guide the compiler to generate the correct code for task based parallelism is of main interest here. In many cases existing sequential functions can be decorated and, with minimal modifications to the code, programmers can direct how these functions will execute and any dependencies at a high level. The types themselves, whilst interesting in their own right, are mainly used by us as a vehicle for illustrating the benefits of driving parallelism through types. We are developing additional types to control task placement, runtime scheduling priorities and resilience.

\end{document}